\documentstyle[preprint,aps]{revtex}
\makeatletter
\@definecounter{whichequation}
\def\themanyeqn{\theequation-\alph{whichequation}}
\def\@manyeqnnum{{\rm (\themanyeqn)}}

\def\manyeqns{\stepcounter{equation}\let\@currentlabel=\themanyeqn
 \setcounter{whichequation}{1}
\global\@eqnswtrue
\global\@eqcnt\z@\tabskip\@centering\let\\=\@manyeqncr
$$\halign to \displaywidth\bgroup\@eqnsel\hskip\@centering
  $\displaystyle\tabskip\z@
 \let\@currentlabel=\theequation
  {##}$&\global\@eqcnt\@ne
  \hfil$\;{##}\;$\hfil
  &\global\@eqcnt\tw@ $\displaystyle\tabskip\z@{##}$\hfil
   \tabskip\@centering&\llap{##}\tabskip\z@\cr}

\def\endmanyeqns{\@@manyeqncr\egroup
    \global\advance\c@equation\m@ne$$\global\@ignoretrue
    \stepcounter{equation}}

\def\@manyeqncr{{\ifnum0=`}\fi\@ifstar{\global\@eqpen\@M
    \@ymanyeqncr}{\global\@eqpen\interdisplaylinepenalty \@ymanyeqncr}}

\def\@ymanyeqncr{\@ifnextchar [{\@xmanyeqncr}{\@xmanyeqncr[\z@]}}

\def\@xmanyeqncr[#1]{\ifnum0=`{\fi}\@@manyeqncr
   \noalign{\penalty\@eqpen\vskip\jot\vskip #1\relax}}

\def\@@manyeqncr{\let\@tempa\relax
    \ifcase\@eqcnt \def\@tempa{& & &}\or \def\@tempa{& &}
      \else \def\@tempa{&}\fi
 \@tempa \if@eqnsw\@manyeqnnum\stepcounter{whichequation}\fi
     \global\@eqnswtrue\global\@eqcnt\z@\cr}
\newcommand{\Cs}[2]{#1_{#2\sigma}^{\dagger}}
\newcommand{\As}[2]{#1_{#2\sigma}}

\newcommand{\Au}[2]{#1_{#2\uparrow}}

\newcommand{\Ad}[2]{#1_{#2\downarrow}}

\def\@cite#1#2{${\hbox{#1\if@tempswa , #2\fi}}$}

\def\Precite{$\hbox{[}$}
\def\Endcite{$\hbox{]}$}

\newcommand{\Eq}[1]{(\ref{eq:#1})}
\newcommand{\eqis}[1]{\label{eq:#1}}
\makeatother

\begin{document}
\baselineskip=18pt 
\title{N\'eel and singlet RVB orders in the {\it t-J} model}
\author{M. Inaba, H. Matsukawa and M. Saitoh}
\address{1$-$16, Department of Physics, Graduate School of Science, Osaka University, Toyonaka, Osaka 560, Japan}

\author{H. Fukuyama}
\address{Department of Physics, University of Tokyo, 7$-$3$-$1 Hongo, Tokyo 113, Japan}
\maketitle
\begin{abstract}
\baselineskip=18pt 
The N\'eel and the singlet RVB orders of the {\it t-J} model in a 2D square lattice are studied in the slave-boson mean-field approximation.
It is shown that the N\'eel order parameter takes the maximum value at the finite temperature and disappear at the lower temperature for a certain range of doping.
It is also shown that the N\'eel and the singlet RVB orders coexist at low temperature.
This suggests the possibility of the coexistence of the N\'eel and the superconducting orders.\\[1.5ex]
\noindent
\end{abstract}
\section{Introduction}
In order to make clear the mechanism of high-$T_{\rm c}$ superconductivity in cuprates,
it is indispensable to understand the whole phase diagram on the temperature-doping plane.
Recent theoretical studies by the slave-boson mean-field theory based on the {\it t-J} model show the validities of the method and the model for high-$T_{\rm c}$ cuprates \Precite\cite{prervb}$-$\nocite{suzumura,tkf}\cite{spinfluc1}\Endcite.
These studies, however, do not take into account the N\'eel order explicitly.
A few works based on the same theoretical scheme study the N\'eel order \Precite\cite{inui,tjmatsu}\Endcite.
In ref. $\!$\cite{inui} the N\'eel order and the singlet RVB order parameters are considered, but the bond order parameters are not treated in the self-consistent way.\quad
In ref. $\!$\cite{tjmatsu} the N\'eel order and the bond order parameters of spinon and holon are included.
It is shown that there is a phase transition between itinerant and localized antiferromagnetism in the N\'eel state.
But the singlet RVB order is not studied there.

In this paper all of the order parameters, {\it i.e. }, the N\'eel order, the RVB order and the bond orders of spinons and holons are considered to study the whole phase diagram in the temperature-doping plane.
It is found that the N\'eel and singlet RVB orders can coexist in the low temperature range.
It is also found that the N\'eel order parameter takes the maximum value at the finite temperature and disappear at the lower temperature for a certain doping region.
\section{Model and Results}
In this communication we study the {\it t-J} model, which is given by
\begin{eqnarray}
 H &=& -t \sum_{<i,j>,\sigma} \left( \tilde{c}_{i\sigma}^{\dagger} \tilde{c}_{j\sigma} + \tilde{c}_{j\sigma}^{\dagger} \tilde{c}_{i\sigma} \right) 
        + J \sum_{<i,j>} {\bf S}_i \cdot {\bf S}_j  \eqis{tj2} \ .
\end{eqnarray}
Here $\tilde{c}_{i\sigma} = c_{i\sigma}(1-n_{i,-\sigma})$, where $\As{c}{i}$ is the electron annihilation operator at site $i$ with spin $\sigma$, ${\bf S}_i$ is the spin operator at site $i$,
$t$ is the transfer integral, $J$ is the exchange coupling constant
and the summation with respect to $<i,j>$ is taken over the nearest neighbor pairs.
We employ the slave-boson method to solve Hamiltonian \Eq{tj2},
in which $\tilde{c}_{i\sigma}$ is replaced by the product of the spinon (fermion) and holon (boson) operators, $\As{f}{i}$ and $b_i$: $\tilde{c}_{i\sigma}=\As{f}{i} b_i^\dagger$.
The spinon and holon numbers should satisfy the local constraint,
\begin{eqnarray}
\sum_{\sigma} \Cs{f}{i} \As{f}{i} +b_i^\dagger b_i=1 \eqis{lconst}.
\end{eqnarray}
The form of the {\it t-J} model eq. $\!$\Eq{tj2} is different from that which is derived from the single-band Hubbard model in the large U limit.
In the latter case the exchange term is $J \sum ({\bf S}_i \cdot {\bf S}_j -(1/4)n_i n_j)$.
The both {\it t-J} model have a local SU(2) symmetry at half filling \Precite\cite{su2}\Endcite.
If we approximate the local constraint \Eq{lconst} to the global one, the latter {\it t-J} model might lose the symmetry.
That is, however, preserved even in that approximation in the present case.
There is another reason of omitting $-(1/4)n_i n_j$ in the $J$-term.
In the case of hole doped high-$T_{\rm c}$ cuprates it is considered that the original microscopic model is the {\it d-p} model.
The {\it t-J} model is derived from the {\it d-p} model through the construction of the Zhang-Rice singlet \Precite\cite{zhang}\Endcite.
In this case the magnitude of the coefficient in the $-n_i n_j$ term is different from 1/4 and is estimated to be negligibly small \Precite\cite{matsupre}\Endcite.

We treat the Hamiltonian in the mean-field approximation, where the local constraint is relaxed to the global one: $\sum_i(\sum_{\sigma} \Cs{f}{i} \As{f}{i} +b_i^\dagger b_i)=N$, where $N$ is the number of the lattice sites.
In order to consider the N\'eel order,
we divide the square lattice into two sublattices A and B and define the N\'eel order parameter (staggered magnetization) $m_{\rm s}=\langle S_A^z\rangle=-\langle S_B^z\rangle$.
We also take into account the uniform bond order parameters $\langle\sum_\sigma \Cs{f}{i}\As{f}{j}\rangle$ and $\langle b_i^\dagger b_j\rangle$, which represent the coherent motions of spinons and holons, respectively,
and the singlet RVB order parameter $\langle\Ad{f}{i}\Au{f}{i+(1,0)}\rangle$ and $\langle\Ad{f}{i}\Au{f}{i+(0,1)}\rangle$, which represents the singlet pairing of spinons.
We assume the $d$-wave pairing: $\langle\Ad{f}{i}\Au{f}{i+(1,0)}\rangle = -\langle\Ad{f}{i}\Au{f}{i+(0,1)}\rangle$,
since the singlet RVB susceptibility for the $d$-wave pairing is shown to diverge at the temperature higher than that for the $s$-wave pairing.

The mean-field Hamiltonian is obtained as
\begin{eqnarray}
 H_{\rm MF} &=&\sum_{k\sigma} {}' \biggl\{ (\epsilon_k-\mu) f_{k\sigma}^\dagger f_{k\sigma}
   + (\epsilon_{k+Q}-\mu) f_{k+Q\sigma}^\dagger f_{k+Q\sigma} \nonumber \\
  & &\makebox[0.5in]{} - \sigma J m_{\rm s} (f_{k\sigma}^\dagger f_{k+Q\sigma} 
               +f_{k+Q\sigma}^\dagger f_{k\sigma} ) \biggl\} \nonumber \\
  &+& \sum_k {}'\ \frac{3}{4}\Bigl(\Delta_k f_{-k\downarrow}^\dagger f_{k\uparrow}^\dagger+ \Delta_{k+Q} f_{-k-Q\downarrow}^\dagger f_{k+Q\uparrow}^\dagger +h.c.\Bigr) \nonumber \\
 &+&\sum_k {}'\ \Bigl((\omega_k - \lambda) b_k^\dagger b_k +(\omega_{k+Q} - \lambda) b_{k+Q}^\dagger b_{k+Q} \Bigr) \nonumber \\
 &+& E_0\ ,
\end{eqnarray}
where 
\begin{eqnarray}
 \epsilon_k &=&-\Big(\ \frac{3}{8} J \sum_\sigma \langle\Cs{f}{i}\As{f}{j}\rangle + t \langle b_i^\dagger b_j\rangle\ \Big) \gamma_k^s \ , \\
 \omega_k &=&-t \sum_\sigma \langle\Cs{f}{i}\As{f}{j}\rangle \gamma_k^s \ ,\\
 E_0&=&4tN\Delta_f \Delta_b +\frac{1}{2} JN m_{\rm s}^2 +\frac{3}{4} JN \Delta_b ^2 
 + 3 JN \Delta_x^2 +\lambda N \ ,
\end{eqnarray}
and $\lambda$ is Lagrange multiplier, which corresponds a holon chemical potential, $\mu=\mu_f+\lambda$ is the spinon chemical potential,where $\mu_f$ is the electron chemical potential, $Q=(\pi,\pi)$, $\gamma_k^s=2(\cos k_x +\cos k_y)$,  $\Delta_k =J\langle\Ad{f}{i}\Au{f}{i+(1,0)}\rangle\gamma_k^d$, $\gamma_k^d=2(\cos k_x -\cos k_y)$,
and the prime on the summation symbol indicates that the sum is taken over the Brillouin zone for the sublattice.
The diagonalized Hamiltonian has four energy bands, $\pm E_{1k}$ and $\pm E_{2k}$,
where $E_{1k}=\sqrt{(\xi_k-\mu)^2 + (3 \Delta_k /4)^2}$ , $E_{2k}=\sqrt{(\xi_k+\mu)^2 + (3 \Delta_k /4)^2}$ and $\xi_k=\sqrt{\epsilon_k^2 + (J m_{\rm s})^2}$ and $\mu < 0$ for $\delta \neq 0$.
The self-consistent equations are given by
\begin{manyeqns}
\delta &=&\frac{1}{N} \sum_k {}' \Biggl[ \frac{1}{\exp((\omega_k - \lambda) /T) -1} 
 + \frac{1}{\exp((\omega_{k+Q} - \lambda) /T) -1} \Biggr]  \eqis{fm} \\[0.07in]
&=& \frac{1}{N} \sum_k {}' \Biggl[ \frac{\xi_k-\mu}{E_{1k}} \tanh \frac{E_{1k}}{2T} 
- \frac{\xi_k+\mu}{E_{2k}} \tanh \frac{E_{2k}}{2T} \Biggr] \ ,\eqis{fl} 
\end{manyeqns}
\begin{eqnarray}
& &\langle b_i^\dagger b_j\rangle
= \frac{1}{4N} \sum_k {}' \Biggl[ \frac{\gamma_k^s}{\exp((\omega_k - \lambda) /T) -1}
 +\! \frac{\gamma_{k+Q}^s}{\exp((\omega_{k+Q} - \lambda) /T) -1} \Biggr] \eqis{ff} \ ,\\[0.07in]
& &\sum_\sigma \langle\Cs{f}{i}\As{f}{j}\rangle
= \frac{-1}{4N} \sum_k {}' \gamma_k^s \frac{\epsilon_k}{\xi_k} \Biggl[ \frac{\xi_k - \mu}{E_{1k}} \tanh \frac{E_{1k}}{2T}
+ \frac{\xi_k + \mu}{E_{2k}} \tanh \frac{E_{2k}}{2T} \Biggr] \ ,\eqis{fb}  \\[0.07in]
& &m_{\rm s}
= \frac{J}{N} \sum_k {}' \frac{m_{\rm s}}{\xi_k} \Biggl[ \frac{\xi_k - \mu}{E_{1k}} \tanh\! \frac{E_{1k}}{2T}
+ \frac{\xi_k + \mu}{E_{2k}} \tanh\! \frac{E_{2k}}{2T} \Biggr] \ ,\eqis{fs} \\[0.07in]
& &\langle\Ad{f}{i}\Au{f}{i+(1,0)}\rangle
= \frac{3}{32N} \sum_k {}' \Delta_k \gamma_k^d \Biggl[ \frac{1}{E_{1k}} \tanh\! \frac{E_{1k}}{2T}
+ \frac{1}{E_{2k}} \tanh\! \frac{E_{2k}}{2T} \Biggr] \ ,\eqis{fr} 
\end{eqnarray}
The above equations are solved numerically for $t/J=4.0$, which is considered to be appropriate for high $T_{\rm c}$ cuprates.

The resultant phase diagram is in Fig. $\!1$.
Here $T_{\rm D}$ (dotted line), $T_{\rm N}$(solid line) and $T_{\rm RVB}$(broken line) are the critical temperatures of the bond order parameters,
 the N\'eel order parameter and the singlet RVB order parameter, respectively,
and in the meshed and hatched regions each order parameter is finite.
Both of the two bond order parameters $\langle\sum_\sigma \Cs{f}{i}\As{f}{j}\rangle$ and $\langle b_i^\dagger b_j\rangle$ become finite at the same temperature $T_{\rm D}$
above which the holons and spinons are localized and below which they are itinerant.
In the left up hatched region enclosed by $T_{\rm N}$ line the N\'eel order exists.
The RVB order survives below the $T_{\rm RVB}$ line.

In order to present the physical pictures for regions with different doping, we divide the $T$-$\delta$ plane into four regions of different $\delta$.
In the region of $\delta$ smaller than $\delta_1$ at which $T_{\rm N}$ equals $T_{\rm D}$, there are the two kind of antiferromagnetism below $T_{\rm N}$.
In the temperature  $T_{\rm N}>T>T_{\rm D}$, localized antiferromagnetism occurs which is due to localized spin.
Below $T_{\rm D}$ itinerant antiferromagnetism occurs.
At $T_{\rm D}$ there is the phase transition between localized and itinerant antiferromagnetism.
At $T_{\rm RVB} < T_{\rm D}$ the singlet RVB order coexists with the N\'eel order.
It is to be noted that $T_{\rm RVB}$ as well as $T_{\rm D}$ go to zero for $\delta=0$.
In the region between $\delta_1$ and $\delta_{\rm c}$,
spinons and holons become itinerant below $T_{\rm D}$ and itinerant antiferromagnetism occurs at $T_{\rm N} < T_{\rm D}$.
The singlet RVB order coexists with the N\'eel order below $T_{\rm RVB} < T_{\rm N}$.
The region between $\delta_c$ and $\delta_2$ is quite interesting.
In this region the N\'eel order parameter exists only in the finite temperature range.
The physical reason for this behavior will be discussed in the next section.
At higher doping rate $\delta >\delta_2$, $T_{\rm RVB}$ decreases monotonically with $\delta$.
At $\delta > 0.275$ $T_{\rm RVB}$ is lower than $10^{-3}$.
In principle, $T_{\rm RVB}$ is finite for finite spinon density.
We could not determine the finite value of $T_{\rm RVB}$ for the sake of numerical accuracy.

The temperature dependence of the staggered magnetization $m_{\rm s}$ is shown in Fig.$\,2$ for various doping rates.
The order parameter takes the maximum value,
{\it i.e.} , the hump at $0<T<T_{\rm N}$ .
This hump is particularly noteworthy for high doping.
For the doping rate $\delta_{\rm c} < \delta < \delta_{\rm 2}$, the order parameter, which increases once below $T_{\rm N}$, vanishes at the finite temperature.
The N\'eel order parameter for a fixed temperature decreases as the doping increases.
\section{Discussion and Summary}

We have studied the {\it t-J} model on the 2D square lattice in the mean-field approximation with the slave-boson description.
We have taken account of four order parameters, $viz$. , the N\'eel order parameter, the d-wave singlet RVB order parameter and the bond order parameters of spinons and holons, and have obtained the phase diagram in the {\it T-$\delta$} plane.

Our result of the temperature dependence of the staggered magnetization $m_{\rm s}$ in Fig. $\!$2 [\cite{chyu}] agrees qualitatively with the neutron scattering experiments of $\rm L_2CuO_{4+\delta}$ by Keimer {\it et. al.} \Precite\cite{expneel}\Endcite.
The staggered magnetization curve against temperature has pronounced peak for samples with lower N\'eel temperature \Precite\cite{chyu2}\Endcite.
In their experiments the doping rate of each samples (Fig. $\!2$(a) in their paper) is not clear,
but the sample with lower N\'eel temperature is considered to be the sample with the larger doping rate.
Therefore their experimental findings is consistent with our result Fig. $\!2$, where the peak of the staggered magnetization is pronounced for larger doping rate.

The physical reason why the staggered magnetization is reduced at the low temperature side is the following.
In general the nesting of Fermi surface with fixed wave vector $Q=(\pi,\pi)$ is hindered by the doping,
since the Fermi surface is shifted from that at half filling, and susceptibility for the $Q$ is reduced.
This reduction is pronounced at lower temperature since at the higher temperature the effect of the shift of the Fermi surface from that at half filling is smeared out\Precite\cite{tkf,spinfluc1}\Endcite.

In Fig. $\!1$ the N\'eel order coexists with the singlet RVB order for the low temperature side.
This suggests the coexistence of the the N\'eel order and the superconducting order.
The superconducting order is characterized by the coexistence of the singlet RVB state and the bose condensation of holons
because of $\langle \Ad{c}{i}\Au{c}{j}\rangle =$ $\langle \Ad{f}{i}\Au{f}{j}\rangle \langle b_i^\dagger b_j^\dagger\rangle $.
The latter does not occur in the exact 2D system at finite temperature,
but may occur
if 3D effect is taken into account in the holon dispersion.
Then the superconducting order can coexist with the N\'eel order in the coexistent phase of the N\'eel and singlet RVB orders.
This coexistent phase has not been observed in the experiment.
The reason is considered that the localization effect induced by disorder associated with doping is much enhanced in the small doping region in the N\'eel order phase, because the effective bandwidth and the carrier number are quite small there.
We expect the experimental observation of the coexistence phase in samples where the disorder effect is small.
For example ${\rm La_2CuO_{4+\delta}}$ may be a good candidate since the metallic state with the N\'eel order is observed and the effect of disorder may be weak.
The preparation of the sample with single phase are desired.

In summary, we have studied the 2D {\it t-J} model in the slave-boson mean-field approximation with four order parameters: the N\'eel, RVB, and bond orders and the phase diagram has been obtained.
It has been shown that the N\'eel and the singlet RVB orders can coexist at low temperature.
We also show that the staggered magnetization takes the maximum value at the finite temperature.
The interesting reentrant behavior of the staggered magnetization is predicted for a certain range of $\delta$.

\clearpage

\clearpage
\section*{Figure Captions}
\makeatletter
\@afterindentfalse
\makeatother
Fig. $\!1$.\hspace{0.2in}  The $\delta$-dependence of the critical temperatures for $t/J$=4.0.
$T_{\rm D}$(dotted line), $T_{\rm RVB}$(broken line) and $T_{\rm N}$(solid line) are the critical temperatures of the bond order of spinon and holon, the RVB order and N\'eel order parameter respectively,
and in the meshed and hatched regions each order parameter becomes non-zero.
In the overlapping regions, the order parameters coexist.
At $\delta_1$ $(\simeq 0.02)$ $T_{\rm D}$ equals $T_{\rm N}$. 
At $\delta_c$ $(\simeq 0.15)$ $T_{\rm N}$=0.
$\delta_2$ $(\simeq 0.16)$ is the maximum doping rate where the N\'eel order exists.\\[0.4in]
Fig. $\!2$.\hspace{0.2in}  The temperature dependence of the N\'eel order parameter, $m_{\rm N}$, at various doping rates.

\begin{thebibliography}{10}
\bibitem{chyu}
Similar behavior is already reported in ref. $\!$\cite{tjmatsu}.
\end{thebibliography}
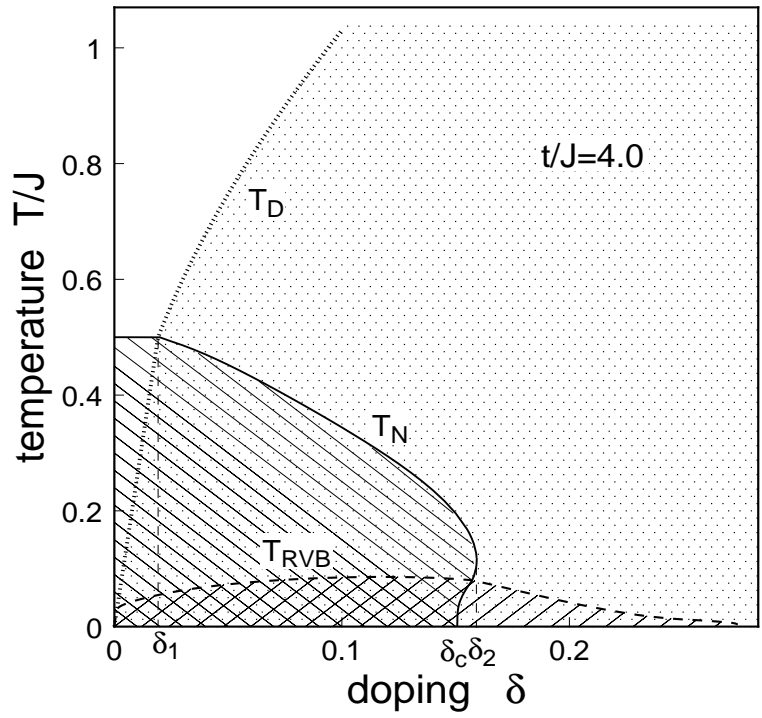
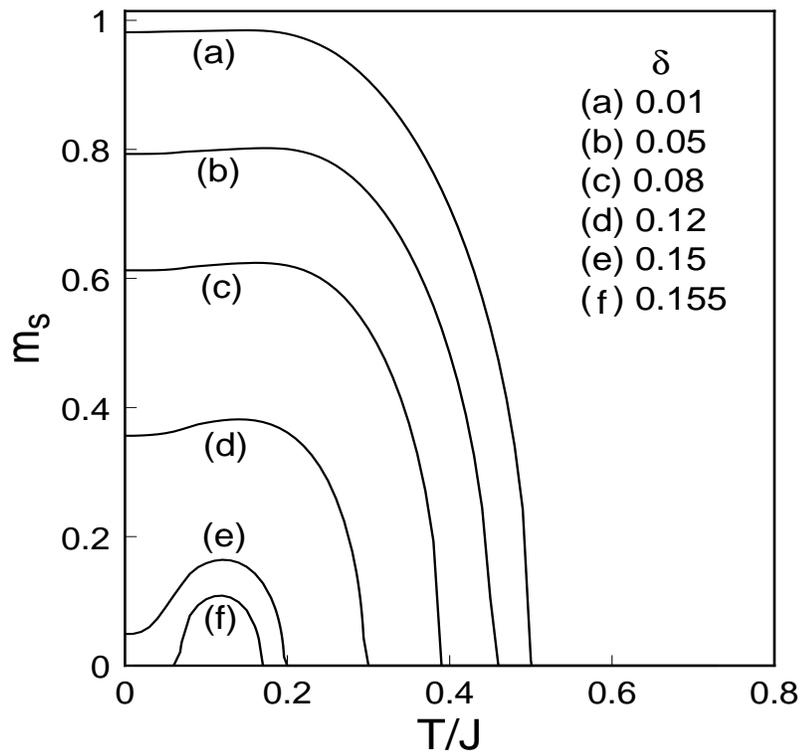
\end{document}